\documentclass[prl,aps,showpacs,twocolumn,unsortedaddress,superscriptaddress]{revtex4-1}
\usepackage{graphics, bm, psfrag, amsmath, amssymb, epsfig, grffile, float}
\usepackage{subfigure}
\usepackage{color}

\begin{document}

\title{Effect of a spin-active interface on proximity-induced superconductivity in topological insulators}

\author{Christopher Triola}
\author{E. Rossi}
\affiliation{Department of Physics, College of William and Mary, Williamsburg, Virginia 23187, USA}
\author{Alexander V. Balatsky}
\affiliation{Institute for Materials Science, Los Alamos, New Mexico 87545, USA}
\affiliation{Nordic Institute for Theoretical Physics (NORDITA), Roslagstullsbacken 23, S-106 91 Stockholm, Sweden}
\date{\today}


\begin{abstract}

We examine the effect of a spin-active interface on the symmetry of proximity-induced superconducting pairing amplitudes in topological insulators. We develop a model to investigate the leading order contribution to the pairing amplitude considering three different kinds of spin-active interfaces: (i) those that induce spin-dependent scattering phases, (ii) those that flip the spin of incident electrons, and (iii) interfaces that both induce spin-dependent phases and flip the spins of incident electrons. We find that in cases (i) and (iii) odd-frequency triplet pairing is induced in the TI while for case (ii) no odd-frequency pairing is induced to leading order. We compare our results to those for normal metals and ferromagnetic materials finding that the nontrivial spin structure of the TI leads to qualitatively different behavior.

\end{abstract}


\maketitle


Strong three dimensional topological insulators (TIs) are a unique class of materials with a
bulk band gap and surfaces hosting topologically protected states
whose momentum and spin degrees of freedom are locked
\cite{ZhangNatPhys2009,HasanRMP2010}. Some examples of
materials now known to be 3D topological insulators are
$\text{Bi}_2\text{Se}_3$, $\text{Bi}_2\text{Te}_3$, and
$\text{Sb}_2\text{Te}_3$ \cite{Chen2009,Xia2009,Hsieh2009}.
The unique electronic properties of TIs make them extremely
interesting for fundamental reasons and for possible uses
in technological applications.
Recent advances in fabrication techniques
allow the realization of heterostructures
with unprecedented control of the thickness, and number,
of layers
\cite{haigh2012}.
The combination of layers of different materials, such as TIs,
superconductors (SCs), graphene, and bilayer graphene
\cite{wang2012, dang2010,song2010}
allows the realization of new systems with novel and extremely
interesting electronic properties
\cite{StanescuPRB2010,YokoyamaPRB2012,Black-SchafferPRB2012,Black-SchafferPRB2013,jzhang2013,JinPRB2013,ZhangTriola2013}.
In particular, it has been shown that Majorana excitations may arise in certain TI/SC heterostructures by including ferromagnetic materials \cite{FuPRL2008,TanakaPRL2009,LinderPRL2010,LinderPRB2010_1}.
Additionally, it has been shown theoretically, and there is experimental evidence to suggest, that in heterostructures formed by a TI and an s-wave SC,
via the proximity effect, p-wave triplet superconducting pairings can
be induced in the TI's surface
\cite{StanescuPRB2010,koren2013}.
More recently it has also been shown that the proximity of a SC to a TI can induce
odd-frequency superconducting pairing in the TI's surface
\cite{YokoyamaPRB2012,Black-SchafferPRB2012,Black-SchafferPRB2013}.

The symmetry of a superconducting state is characterized by
the symmetry properties of the pairing amplitude
$F(\textbf{r}_1,t_1;\textbf{r}_2,t_2)=\sum_{\alpha,\beta}\left< T
c_\alpha(\textbf{r}_1,t_1)c_\beta(\textbf{r}_2,t_2)
\right>g_{\alpha\beta}$, where $g_{\alpha\beta}$ is a metric tensor
describing the spin structure of the pair. Because electrons are
fermions if $g_{\alpha\beta}$ describes a spin singlet then the equal
time correlation function must be even in parity
$F(\textbf{r}_1,t;\textbf{r}_2,t)=F(\textbf{r}_2,t;\textbf{r}_1,t)$
and if it describes a spin triplet then the equal time correlation
function must be odd in parity
$F(\textbf{r}_1,t;\textbf{r}_2,t)=-F(\textbf{r}_2,t;\textbf{r}_1,t)$. However,
spin triplet pairs can be even in parity and spin singlet pairs can be
odd in parity if the pairing amplitude is odd in time or, equivalently, Matsubara
frequency, as was originally proposed for superfluid $\text{He}^3$
\cite{Berezinskii1974} and later for superconductivity
\cite{BalatskyPRB1992}. This ensures that equal time correlations
vanish enforcing the Pauli principle and leads to a rich variety of
pairing symmetries.
Odd-frequency pairing has been shown to develop in
ferromagnetic insulator/superconductor (FMI$\mid$SC)
\cite{EschrigNat2008}, ferromagnetic metal/superconductor (FMM$\mid$SC)
\cite{LinderPRB2008}, and normal metal/superconductor (N$\mid$SC) junctions \cite{TanakaPRB2007,LinderPRL2009,LinderPRB2010_2,TanakaJPSJ2012}.
Several of these works \cite{EschrigNat2008,LinderPRB2008,LinderPRL2009,LinderPRB2010_2} obtained the proximity-induced odd-frequency pairing amplitudes by including the effect of a {\em spin-active} interface, i.e. an interface that induces a spin dependence of the transmission and reflection amplitudes of the fermionic quasiparticles. These works found that a spin-active interface can modify qualitatively the nature
of the pairing amplitude in N$\mid$SC, FMI$\mid$SC, and FMM$\mid$SC heterostructures.

\begin{figure}[H]
 \begin{center}
  \centering
  \includegraphics[width=4.0cm]{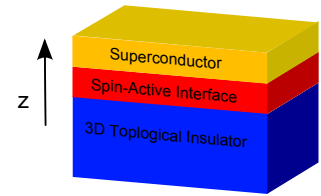}
  \caption{
           (Color online) Sketch of the TI$\mid$SC heterostructure considered.
           spin-active interface is present between the superconductor and the 3D topological insulator.
           The spin-active interface could be realized by a thin layer of magnetic material such as EuO.}
  \label{fig:TISC}
 \end{center}
\end{figure}
In this work we investigate the effect of a
spin-active interface on the symmetry of the superconducting pairing
induced in the TI's surface by the proximity of an s-wave
superconductor. Previous works on TI$\mid$SC heterostructures \cite{StanescuPRB2010,YokoyamaPRB2012,Black-SchafferPRB2012,Black-SchafferPRB2013}
had not taken into account the presence of a spin-active interface.
In principle any interface between two materials whose quasiparticle spin states are different can be thought of as spin-active. However, one could also engineer an interface,	A$\mid$B, to be spin-active by inserting a thin layer of magnetic material between A and B. Below we develop a model to describe a generic spin-active interface between two effectively 2D systems. We then apply it to the case of a TI$\mid$SC heterostructure with a spin-active interface. Our results show that the presence of a spin-active interface profoundly
affects the nature of the proximity-induced superconducting pairing in the TI.
In particular, we find that in TI$\mid$SC heterostructures with a spin-active
interface the odd-frequency components of the pairing
amplitude have different spin and spatial structure from the ones
of TI$\mid$SC heterostructures with no spin-active interface and
from the ones of N$\mid$SC, FMI$\mid$SC, and FMM$\mid$SC
heterostructures with spin-active interfaces
\cite{TanakaPRB2007,EschrigNat2008,LinderPRB2008,LinderPRL2009,LinderPRB2010_2,TanakaJPSJ2012}.

Figure~\ref{fig:TISC} shows schematically a TI-SC heterostructure
with a spin-active interface. We
consider three kinds of spin-active interfaces: those which confer a spin-dependent interfacial phase (SDIP) to quasiparticle states at the
interface; those that flip the spins of quasiparticles at the interface; and those that do
both.
By SDIPs we refer to the process whereby quasiparticle states
incident on the interface pick up a phase when transmitted
$\left| \uparrow \right>_{\textbf{k}}\rightarrow e^{i\theta_{\uparrow,\textbf{k}}}\left| \uparrow\right>_{\textbf{k}}$ and $\left| \downarrow \right>_{\textbf{k}}\rightarrow e^{i\theta_{\downarrow,\textbf{k}}}\left| \downarrow \right>_{\textbf{k}}$. The spin and $\textbf{k}$-dependence of the phases $\theta_{\alpha,\textbf{k}}$ are determined by the microscopic details of the interface \cite{MillisPRB1988,EschrigPRL2003,CottetPRB2009,LinderPRB2010_2}. This process is thought to be a common feature of spin-active interfaces \cite{EschrigNat2008,EschrigPRL2003,TokuyasuPRB1988,CottetPRB2005,KopuPRB2004,BergeretPRL2001} and can be thought of as a precession of the incident electron's spin about the
magnetization axis of the interface. Let $\eta_\textbf{k}\equiv (\theta_{\uparrow,\textbf{k}} + \theta_{\downarrow,\textbf{k}}+\theta_{\uparrow,-\textbf{k}} + \theta_{\downarrow,-\textbf{k}})/2$, $\delta\theta_\textbf{k}\equiv \theta_{\uparrow,\textbf{k}} - \theta_{\downarrow,\textbf{k}}$ and $\zeta_\textbf{k}\equiv (\delta\theta_{\textbf{k}} - \delta\theta_{-\textbf{k}})/2$, using this convention a spin-singlet pair $\left| \uparrow\right>_\textbf{k}\left| \downarrow\right>_{-\textbf{k}}-\left| \downarrow\right>_\textbf{k}\left| \uparrow\right>_{-\textbf{k}}$ is converted to $e^{i\eta_\textbf{k}}\left(e^{i\zeta_\textbf{k}}\left| \uparrow\right>_\textbf{k}\left| \downarrow\right>_{-\textbf{k}}-e^{-i\zeta_\textbf{k}}\left| \downarrow\right>_\textbf{k}\left| \uparrow\right>_{-\textbf{k}}\right)$ upon scattering at the interface. Hence a singlet pair in the superconductor develops a triplet component proportional to $\sin\zeta_\textbf{k}$ at the interface. Thus we can see that the most important consequence of the presence of SDIPs is the conversion of purely spin singlet
pairing amplitudes to a linear combination of singlet and triplet amplitudes at the interface. Any material that possesses this property could be used to capture the effects we derive for SDIPs.
By spin-flipping (SF) we refer to tunneling processes that
do not conserve the spin of transmitted electrons. This
process could be realized by any material whose quasiparticle
states are in a spin state that is a different linear
combination of spin up and spin down from the
superconductor. An example of this kind of material would be a
ferromagnetic half-metal.

The main difference between a topological insulator and other materials for which the effect of spin-active interfaces have been studied is that, at low energies, topological insulator states possess a spin lying in the plane of the surface whose direction is locked with the direction of the momentum. We will show that this affects the symmetries of the induced pairing, creating odd-frequency m = ±1 triplet (S = 1; m = ±1) correlations for any spin-active interface that confers SDIPs.

To model the system in Fig \ref{fig:TISC} we employ the Hamiltonian: $H = H_{TI} + H_{SC} + H_t$ where:
\begin{equation}
\begin{aligned}
H_{TI}&=\sum_{\textbf{k},\lambda,\lambda'}(\hbar v \hat{z}\cdot\boldsymbol\sigma\times\textbf{k}-\mu\sigma_0)_{\lambda \lambda'}c^\dagger_{\textbf{k},\lambda}c_{\textbf{k},\lambda'}  \\
H_{SC}&=\sum_{\textbf{k},\lambda,\lambda'}\left(\epsilon_\textbf{k}d^\dagger_{\textbf{k},\lambda}d_{\textbf{k},\lambda} + \hat{\Delta}_{\lambda \lambda'}d^\dagger_{\textbf{k},\lambda}d^\dagger_{-\textbf{k},\lambda'}\right) + \text{h.c.}  \\
H_t&=\sum_{\textbf{k},\lambda,\lambda'}\hat{T}_{\lambda \lambda'} c^\dagger_{\textbf{k},\lambda}d_{\textbf{k},\lambda'} + \text{h.c.}
\end{aligned}
\label{eq:h}
\end{equation}
where $\sigma_0$ is the $2\times2$ identity matrix in spin space, $\boldsymbol\sigma$ is the vector $(\sigma_1, \sigma_2, \sigma_3)$ formed by $2\times2$ Pauli matrices in spin space, $\textbf{k}=(k_x,k_y,0)$, $v$ is the Fermi velocity of the surface states in the TI, $\mu$ is the chemical potential in the TI surface, $c^\dagger_{\textbf{k},\lambda}$ $\left(d^\dagger_{\textbf{k},\lambda}\right)$ creates a quasiparticle with momentum $\textbf{k}$ and spin $\lambda$ in the TI surface (superconductor), $\epsilon_\textbf{k}$ is the energy of a superconductor quasiparticle state measured from the chemical potential in the superconductor, $\hat{\Delta}=-\Delta_0 i \sigma_2$ is the order parameter of the superconducting condensate, and $\hat{T}=\left(t_0\sigma_0 +\textbf{t}\cdot\boldsymbol\sigma \right)$ with $\textbf{t}=(t_1,t_2,t_3)$. Notice that the tunneling term accounts for the possibility of spin-flip processes at the interface if $\textbf{t}\neq0$.

To investigate the effect of the spin-active interface on proximity-induced pairing in the TI we calculate the pairing amplitude in the TI as a function of momentum $\textbf{k}$ and Matsubara frequency $\omega$, $\hat{F}^{TI}(\textbf{k},\omega)$.
To leading order in $\hat{T}$ we have:
\begin{equation}
\hat{F}^{TI}(\textbf{k},\omega) = \hat{G}^{TI}_0(\textbf{k},\omega)\hat{T}\hat{F}^{SC}_{\theta_\textbf{k}}(\textbf{k},\omega)\hat{T}^\text{T}\hat{G}^{TI}_0(-\textbf{k},-\omega)^\text{T}
\label{eq:ftisc}
\end{equation}
where we have included SDIP by a transformation in spin-space at the interface $\hat{F}^{SC}_{\theta_\textbf{k}}(\textbf{k},\omega)=e^{i\eta_\textbf{k}}e^{i\frac{\delta\theta_{\textbf{k}}}{2}\sigma_3}\hat{F}^{SC}_0(\textbf{k},\omega)e^{i\frac{\delta\theta_{-\textbf{k}}}{2}\sigma_3}$, where $\hat{F}^{SC}_0(\textbf{k},\omega)=-\hat{\Delta}/\left(\omega^2 +\epsilon_\textbf{k}^2 +\Delta_0^2 \right)$
is the pairing amplitude in the SC.

Evaluating the expression on the right hand side of  Eq (\ref{eq:ftisc}) we find $\hat{F}^{TI}(\textbf{k},\omega)=\frac{-i\Delta_0}{(\omega^2 + \epsilon_\textbf{k}^2 + \Delta_0^2)\left[(i\omega+\mu)^2 - \hbar^2v^2k^2\right]\left[(i\omega-\mu)^2 - \hbar^2v^2k^2\right]}e^{i\eta_\textbf{k}}\hat{f}^{TI}(\textbf{k},\omega)$ where
\begin{equation}
\hat{f}^{TI}(\textbf{k},\omega)=f^{TI}_0 \sigma_0 + f^{TI}_1 \sigma_1 + f^{TI}_2 \sigma_2 + f^{TI}_3 \sigma_3
\label{eq:fti}
\end{equation}
and
\begin{widetext}
\begin{equation}
\begin{aligned}
f^{TI}_0&= 2\sin\zeta_\textbf{k}\left[ -\left(\omega^2 + \mu^2 + \hbar^2v^2(k_x^2-k_y^2)\right)\left(t_0t_1-it_2t_3 \right) - 2\hbar^2v^2k_xk_y\left(t_0t_2+it_1t_3\right) + i\omega\hbar v k_y(t_0^2-2t_3^2+|\textbf{t}|^2)\right] \\
&+2\cos\zeta_\textbf{k}\left[ \hbar v k_x\mu(t_0^2-|\textbf{t}|^2) \right] \\
f^{TI}_1&= \sin\zeta_\textbf{k}\left[-\left(\omega^2 + \mu^2 - \hbar^2v^2k^2\right)\left(t_0^2-2t_3^2+|\textbf{t}|^2\right) - 4i\omega\hbar v \left[ k_x(t_0t_2+it_1t_3)-k_y(t_0t_1-it_2t_3) \right] \right] \\
f^{TI}_2&= \sin\zeta_\textbf{k}\left[ 4\mu\hbar v \left[ k_x(t_0t_1-it_2t_3)+k_y(t_0t_2+it_1t_3) \right] \right]-\cos\zeta_\textbf{k}\left[ \left(\omega^2 + \mu^2 + \hbar^2v^2k^2\right)\left(t_0^2-|\textbf{t}|^2\right) \right] \\
f^{TI}_3&= -2\sin\zeta_\textbf{k}\left[ \left(\omega^2 + \mu^2 - \hbar^2v^2(k_x^2-k_y^2)\right)\left(t_1t_3-it_0t_2 \right) - 2\hbar^2v^2k_xk_y\left(t_2t_3+it_0t_1\right) + \omega\hbar v k_x(t_0^2-2t_3^2+|\textbf{t}|^2) \right] \\
&-2\cos\zeta_\textbf{k}\left[ i\hbar v k_y\mu(t_0^2-|\textbf{t}|^2) \right].
\end{aligned}
\label{eq:fti0123}
\end{equation}
\end{widetext}
The $S=1$ $m=\pm1$ components of the pairing amplitude are given by $f^{TI}_0\pm f^{TI}_3$, the $m=0$ triplet component by $f^{TI}_1$, while the singlet ($S=0$) is given by $f^{TI}_2$. From  Eq (\ref{eq:fti0123}) we can see that the presence of a spin-active interface induces odd-frequency triplet correlations in the TI, similar to the case where the TI layer is replaced by a 3D normal metal or ferromagnetic material \cite{DemlerPRB1997,BergeretPRL2001,EschrigPRL2003,CottetPRB2005,LinderPRB2008,LinderPRL2009,LinderPRB2010_2}.
It is interesting to note that the $m=\pm 1$ amplitudes possess a non-trivial $\textbf{k}$-dependence reminiscent of a chiral state. Specifically, the odd-frequency components are proportional to $|\textbf{k}|\sin\zeta_\textbf{k}e^{\mp i\phi_\textbf{k}}$ while the even-frequency components are proportional to $|\textbf{k}|^2\sin\zeta_\textbf{k}e^{\mp i2\phi_\textbf{k}}$ where $\phi_\textbf{k}=\tan^{-1}k_y/k_x$. From Eqs (\ref{eq:fti0123}) we note that if there are no SDIPs, that is $\zeta_{\textbf{k}}=0$, then the $f^{TI}_1$ component does not contribute to $\hat{f}^{TI}(\textbf{k},\omega)$ and the $f^{TI}_0$ and $f^{TI}_3$ components are proportional to $\mu$ so that at the Dirac point no triplet correlations are induced in the TI at this order. The next term contributing to $\hat{F}^{TI}(\textbf{k},\omega)$ is proportional to $\hat{T}^4$ and at this order we do find odd-frequency triplet correlations even with $\zeta_\textbf{k}=0$, however these amplitudes are orders of magnitude smaller than the singlet contribution in Eq (\ref{eq:fti0123}) and will not be presented here.

If instead we have $\zeta_\textbf{k}\neq0$ and no spin-flipping ($\textbf{t}=0$) then Eqs (\ref{eq:fti0123}) simplify to:
\begin{equation}
\begin{aligned}
f^{TI}_0&=  2t_0^2\hbar v \left[\mu\cos\zeta_\textbf{k} k_x + i\omega\sin\zeta_\textbf{k} k_y \right] \\
f^{TI}_1&= -t_0^2\sin\zeta_\textbf{k}\left(\omega^2 +\mu^2 - \hbar^2v^2k^2\right)  \\
f^{TI}_2&= -t_0^2\cos\zeta_\textbf{k}\left(\omega^2 +\mu^2 + \hbar^2v^2k^2\right) \\
f^{TI}_3&= -2t_0^2\hbar v \left[\omega\sin\zeta_\textbf{k} k_x + i\mu\cos\zeta_\textbf{k} k_y \right].
\end{aligned}
\label{eq:fti0123t0}
\end{equation}
From these equations we see that even in the absence of spin-flip processes SDIPs lead to chiral odd-frequency $m=\pm1$ triplet pairing on a TI surface. However, spin-flip processes are necessary to give rise to odd-frequency $m = 0$ triplet pairing and even-frequency $m =\pm 1$ triplet pairing at the Dirac point of the TI.

To gain some insight, we compare these results to the case of a X$\mid$S junction with a spin-active interface where we take X to be a 2D material described by the Hamiltonian $H_X=\sum_{\textbf{k},\lambda}(\xi_\textbf{k}\sigma_0+\textbf{h}\cdot\boldsymbol\sigma)_{\lambda\lambda}a^\dagger_{\textbf{k},\lambda}a_{\textbf{k},\lambda}$ where we assume $\xi_{-\textbf{k}}=\xi_\textbf{k}$. For $\textbf{h}=0$ this describes a 2D normal metal (X=N), for $\textbf{h}\neq 0$ this describes a ferromagnet (X=F). We make a distinction between two limits of the F case, one in which $\textbf{h}=(0,0,h)$ (FZ) and an easy-plane ferromagnet $\textbf{h}=h(\cos\phi,\sin\phi,0)$ (FE). To calculate the leading order contribution to the anomalous Green's function for this kind of system, $\hat{F}^{X}(\textbf{k},\omega)$, (ignoring the effect of the exchange field on the superconductor) we replace $\hat{G}^{TI}_0(\textbf{k},\omega)$ in Eq (\ref{eq:ftisc}) with $\hat{G}^{X}_0(\textbf{k},\omega)=\frac{1}{(\xi_{\textbf{k}}-i\omega)^2-\left|\textbf{h}\right|^2}[(i\omega-\xi_\textbf{k})\sigma_0+\textbf{h}\cdot\boldsymbol\sigma]$. Evaluating the resulting expression we find $\hat{F}^{X}(\textbf{k},\omega)=\frac{-i\Delta_0}{(\omega^2+\Delta^2+\epsilon_\textbf{k}^2)[\xi_\textbf{k}^4+2\xi_\textbf{k}^2(\omega^2-\left|\textbf{h}\right|^2)+(\omega^2+\left|\textbf{h}\right|^2)^2]}e^{i\eta_\textbf{k}}\hat{f}^X(\textbf{k},\omega)$
where
\begin{equation}
\begin{aligned}
\hat{f}^{X}(\textbf{k},\omega)&=f^{X}_0 \sigma_0+f^{X}_1 \sigma_1+f^{X}_2 \sigma_2 +f^{X}_3 \sigma_3. \\
\end{aligned}
\label{eq:fx}
\end{equation}
and
\begin{equation}
\begin{aligned}
f^{X}_0&=-i2\cos\zeta_\textbf{k} \omega h_2 \left( t_0^2 -|\textbf{t}|^2 \right) \\
&- 2\sin\zeta_\textbf{k} \left(h_1\xi_\textbf{k}+ih_2h_3 \right)\left( t_0^2-2t_3^2+|\textbf{t}|^2 \right) \\
&+2\sin\zeta_\textbf{k} \left( \omega^2 + \xi_\textbf{k}^2-2h_2^2 + |\textbf{h}|^2\right)\left(t_0t_1-it_2t_3 \right) \\
&+4\sin\zeta_\textbf{k}\left(h_1h_2 + ih_3\xi_\textbf{k} \right)\left(t_0t_2+it_1t_3 \right)   \\
f^{X}_1&= 2\cos\zeta_\textbf{k} \omega h_3 \left( t_0^2 -|\textbf{t}|^2 \right) \\
&- \sin\zeta_\textbf{k} \left( \omega^2 + \xi_\textbf{k}^2 +2h_1^2 -|\textbf{h}|^2\right)\left(t_0^2-2t_3^2+|\textbf{t}|^2\right) \\
&-4\sin\zeta_\textbf{k} \left( h_2h_3+ih_1\xi_\textbf{k}   \right)\left(t_2t_3+it_0t_1 \right) \\
&+4\sin\zeta_\textbf{k}\left(h_2\xi_\textbf{k} + ih_1h_3 \right)\left(t_0t_2+it_1t_3 \right)  \\
f^{X}_2&= \cos\zeta_\textbf{k} \left( \omega^2 + \xi_\textbf{k}^2 -|\textbf{h}|^2\right) \left( t_0^2 -|\textbf{t}|^2 \right) \\
&-2 \sin\zeta_\textbf{k} \omega\left[ 2h_1\left(t_1t_3-it_0t_2 \right) + 2h_2\left(t_2t_3+it_0t_1 \right)\right] \\
&+2 \sin\zeta_\textbf{k}\omega h_3\left(t_0^2-2t_3^2+|\textbf{t}|^2 \right) \\
f^{X}_3&= 2\cos\zeta_\textbf{k} \omega h_1 \left( t_0^2 -|\textbf{t}|^2 \right) \\
&+ 2\sin\zeta_\textbf{k} \left(h_1h_3+ih_2\xi_\textbf{k} \right)\left( t_0^2-2t_3^2+|\textbf{t}|^2 \right) \\
&+2\sin\zeta_\textbf{k} \left( \omega^2 + \xi_\textbf{k}^2 -2h_1^2 +|\textbf{h}|^2\right)\left(t_1t_3-it_0t_2 \right) \\
&-4\sin\zeta_\textbf{k}\left(h_3\xi_\textbf{k}+ ih_1h_2  \right)\left(t_0t_1-it_2t_3 \right) .
\end{aligned}
\label{eq:fx0123}
\end{equation}

Notice that the odd-frequency $m=0$ triplet component is proportional to $h_3\cos\zeta_\textbf{k}$, while the $m=\pm1$ triplet component is proportional to $(h_2\pm ih_1)\cos\zeta_\textbf{k}$ hence if the material has a non-zero exchange field then even for $\zeta_\textbf{k}=0$ there is an odd-frequency triplet amplitude in contrast to the case of either a normal metal or a TI. At this point we can use the components in Eqs (\ref{eq:fti0123}, \ref{eq:fx0123}) to explore the properties of the cases noted above. The symmetries for the four systems TI$\mid$SC, N$\mid$SC, FZ$\mid$SC, and FE$\mid$SC are summarized in Table \ref{table:results}.

\begin{widetext}
\begin{center}
\begin{table}[htb]
\caption{Comparison of Proximity-Induced Pairing in TI$\mid$SC, N$\mid$SC, FZ$\mid$SC, and FE$\mid$SC}
\begin{tabular}{l | l c c c c}
\hline\hline
 & Interface & TI$\mid$SC & N$\mid$SC & FZ$\mid$SC & FE$\mid$SC  \\ [0.5ex]
\hline
 & No SF or SDIP  & $S=0,1; \ m=\pm1$ & $S=0$ & $S=0$ & $S=0$ \\
Even-$\omega$ & SDIP  & $S=0,1; \ m=0,\pm1$ & $S=0,1; \ m=0$ & $S=0,1; \ m=0$ & $S=0,1; \ m=0,\pm1$  \\
 & SF  & $S=0,1; \ m=\pm1$ & $S=0$ & $S=0$ & $S=0$ \\
 & SF and SDIP  & $S=0,1; \ m=0,\pm1$ & $S=0,1; \ m=0,\pm1$ & $S=0,1; \ m=0,\pm1$ & $S=0,1; \ m=0,\pm1$ \\ [1ex]
\hline
 & No SF or SDIP  & -- & -- & $S=1; \ m=0$ & $S=1; \ m=\pm1$ \\
Odd-$\omega$ & SDIP  & $S=1; \ m=\pm1$ & -- & $S=0,1; \ m=0$ & $S=1; \ m=\pm1$   \\
 & SF  & -- & -- & $S=1; \ m=0$ & $S=1; \ m=\pm1$ \\
 & SF and SDIP  & $S=1; \ m=0,\pm1$ & -- & $S=0,1; \ m=0$ & $S=0,1; \ m=\pm1$ \\ [1ex]
\hline
\end{tabular}
\label{table:results}
\end{table}
\end{center}
\end{widetext}

Table \ref{table:results} shows that the presence of an interface with SDIPs induces odd-frequency triplet correlations in TI$\mid$SC heterostructures. Another feature of Table \ref{table:results} is that the FZ$\mid$SC and N$\mid$SC only develop $m\pm1$ triplet amplitudes if the interface both confers SDIPs and is spin-flipping, in contrast to the TI$\mid$SC and FE$\mid$SC which exhibit $m=\pm1$ triplet amplitudes for all four interfaces. This can be explained by realizing that the SDIPs convert a singlet pair into a linear combination of singlet and m = 0
triplet but this mechanism cannot align two spins in a Cooper pair that were originally anti-aligned. Spin-flipping processes can take the $m=0$ triplet state and rotate it out of the plane to produce an $m=\pm1$ triplet. In the case of the FE and TI, the spin of the eigenstates for these materials lies in the $x$-$y$ plane and hence these states are already a linear combination of $\left|\uparrow \right>$ and $\left|\downarrow \right>$. This acts as an intrinsic mechanism for aligning the spins of the paired quasiparticles. For this reason we can see that the FE and TI exhibit $m=\pm1$ triplet contributions for all four interfaces.

It is worth noting that the symmetries of the induced pairings in the FE are not sensitive to the value of the chemical potential while in the case of the TI, for interfaces that lack SDIPs, the only triplet contributions are proportional to $\mu$ so that at the Dirac point an interface without SDIPs will only give rise to singlet pairing in the TI. Another difference between the TI and FE is that for the TI odd-frequency pairing only develops in the presence of SDIPs while odd-frequency pairing is ubiquitous in the FE (and FZ) for all four interfaces. These qualitative differences between the TI and FE results can be attributed to the chiral spin structure of the TI, i.e. the fact that $\textbf{k}\rightarrow-\textbf{k}$ implies $\textbf{s}\rightarrow-\textbf{s}$, where $\textbf{s}$ is the spin of an electron on the surface of a TI.

Note that for the normal metal we see that no odd-frequency amplitudes are induced at this order. We attribute this to the trivial spin structure of the normal metal whose Green's function is even in frequency and proportional to the identity in spin space so the only way to induce odd-frequency correlations in this material would be through processes of higher order in $\hat{T}$.

In conclusion, we analyzed proximity-induced superconductivity in TI$\mid$S
heterostructures with a spin-active interface. We find that the proximity-induced pairing amplitudes in the TI are qualitatively different from non-chiral materials.
The presence of spin-dependent interfacial phases gives rise to odd-frequency
$m=\pm1$ triplet correlations. This appears to be due to the unique spin structure of the TI
surface states. Another interesting feature of the
$m=\pm1$ triplet correlations for TI$\mid$S structures with a spin active
interface is the fact that both the even and odd-frequency
contributions possess non-trivial $\textbf{k}$-dependence reminiscent of a chiral state, the odd-frequency terms
being proportional to $\sin\zeta_\textbf{k}e^{-i\phi_\textbf{k}}$ while the even-frequency
terms are proportional to $\sin\zeta_\textbf{k}e^{-i2\phi_\textbf{k}}$. Additionally, we
find that the magnitude of the odd-frequency
pairing amplitude is dependent on the direction of
$\textbf{t}$ a quantity that could be tuned by appropriately
manufacturing the interface. Depending on the degree of control one
has on the direction of $\textbf{t}$, this could allow for the ability
to turn the odd-frequency pairing amplitude on or off as desired. This
property could provide a new tool in classifying different materials
experimentally and could find applications in the field of
spintronics.

We are grateful to A. Black-Schaffer for useful discussions. This work was supported by ONR, Grant No. ONR-N00014-13-1-0321 (C.T. and E.R.), ACS-PRF Grant No. 53581-DNI5 (C.T. and E.R.), the Jeffress Memorial Trust (C.T. and E.R.), the Virginia Space Grant Consortium (C.T.), and USDOE, VR, and ERC-DM-32103 (A.B.). C.T. acknowledges the hospitality of LANL.

%

%


\end{document}